\documentclass[showpacs,twocolumn]{revtex4}

\usepackage{graphicx}
\usepackage{dcolumn}
\usepackage{amsmath}

\makeatletter
\def\btt#1{\texttt{\@backslashchar#1}}
\DeclareRobustCommand\bblash{\btt{\@backslashchar}} \makeatother

\begin{document}

\title[]{Rotating black hole in Einstein and pure Lovelock gravity}
\author{Naresh Dadhich$^{a,\;b}$}\email{nkd@iucaa.ernet.in}
\author{Sushant~G.~Ghosh$^{a}$} \email{sghosh2@jmi.ac.in, sgghosh@gmail.com}
\affiliation{$^{a}$ Center for Theoretical Physics, Jamia Millia
Islamia, New Delhi 110025, India}
\date{\today}
\affiliation{$^{b}$ Inter-University Centre for Astronomy \&
Astrophysics, Post Bag 4, Pune 411 007, India}

\begin{abstract}
We obtain rotating black hole metric for higher dimensional Einstein and pure
Lovelock gravity by employing two independent and well motivated methods. One is based on the principle of incorporation of Newtonian acceleration for timelike motion while no acceleration for null motion. The other is the well known Newman-Janis alogrithm that converts a static black hole into a rotating one through a complex transformation. It turns out that both the methods give the same metric which for Einstein black hole is an exact vacuum solution while for pure Lovelock black hole it satisfies the vacuum equation in the leading order. However it shares all the physical properties with the well known Kerr black hole.
\end{abstract}



\maketitle


It is well known that Lovelock gravity is the natural generalization
of Einstein gravity in higher dimensions. It arises as higher order
polynomial invariants constructed from Riemann curvature where the
linear order corresponds to Einstein gravity. It is also known that
this generalization has two remarkable and distinguishing
properties: (a) the equation continues to be second order
notwithstanding polynomial character of Lagrangian and (b) higher
order terms make non-zero contribution in the equation only for
dimensions $d\geq2n+1$, where $n$ is degree of the polynomial. To
these, there has recently been added a universal feature \cite{dgj}
that gravity is always kinematic in odd $d=2n+1$ and it becomes
dynamic in even $d=2n+2$ dimension. This is true for all $n$. For
$n=1$, Einstein gravity is kinematic in $3$ dimension (i.e. $R^{(1)}_{ab} = 0$
implies $R^{(1)}_{abcd}=0$) and becomes dynamic in $4$ dimension. It
is possible to define $n$-th order Riemann curvature,
$R^{(n)}_{abcd}$ with the property that trace of its Bianchi
derivative vanishes and thereby yielding $n$-th order analogue of
Einstein tensor, $G^{(n)}_{ab}$ (the same as what follows from variation of $n$-th
order Lovelock Lagrangian), with vanishing  divergence
\cite{bianchi}. That is, in odd $d=2n+1$ dimension, $G^{(n)}_{ab}=0$
implies $R^{(n)}_{abcd}=0$;i.e. it is Lovelock flat but however not
Riemann flat \cite{dgj}. Hence pure Lovelock gravity is always kinematic in all
odd $d=2n+1$ dimensions and it turns dynamic in even $d=2n+2$ dimension. This is the most remarkable and unique feature of pure Lovelock gravity.

It also turns out that thermodynamical parameters of static black
holes always  bear the same relation with horizon radius \cite{dpp1}
for $d=2n+1, 2n+2$. For instance, entropy would always go as square
of horizon radius which in terms of black hole area goes as $A^{1/n}$
in all even $d=2n+2$ dimension. This is a kind of
thermodynamical universality of pure Lovelock gravity. Further, static
vacuum solutions with $\Lambda$ go  over asymptotically to the
corresponding Einstein Schwarzschild-de Sitter solution even though
the equation is free of Einstein term \cite{dpp2}. This indicates
that pure Lovelock gravity automatically includes Einstein gravity in low
energy regime.

All these considerations strongly point to the fact that right gravitational
equation is pure Lovelock equation which is valid for $d=2n+1, 2n+2$
dimensions.  That is to what order of Riemann curvature should be
included is determined by spacetime dimension? It is therefore
strongly argued that proper gravitational equation in higher
dimension \cite{prague} is
\begin{equation}
G^{(n)}_{ab} = -\Lambda g_{ab} - \kappa T_{ab},
\end{equation}
for $d=2n+1, 2n+2$ dimension. The most attractive feature of this
proposal  is universality of the basic character of gravity in terms
of being kinematic and dynamic in odd and even dimensions
respectively. This is unique to pure Lovelock gravity as neither is
it true for Einstein nor for any of its other generalizations. One
other elementary gravitational property is existence of bound orbits
around a static isolated source. For Einstein gravity (so for Newton
as well), bound orbits can exist only in $d=4$ dimension and in none
else while for pure Lovelock gravity they do for all even $d=2n+2$
dimension \cite{dgj1}.

Higher order terms in Lovelock action are generally taken as high energy
corrections to Einstein gravity. It is motivated by string theoretic
calculations where one loop integration yields Gauss-Bonnet terms alongwith much
else in low energy effective theory \cite{fri}. In contrast to that, we would
like to emphasize that the above equation stands of its own as a gravitational equation in
higher dimensions \cite{prague}. It is not a correction to Einstein gravity but instead it
says that it is a proper equation for all odd and even dimensions what Einstein gravity is in $3$ and $4$ dimensions. That is for a given $n$, the relevant dimension is $d=2n+1, 2n+2$. This is what we mean by pure Lovelock gravity.

With this motivation, it becomes pertinent to probe Lovelock gravity
in  various situations in higher dimensions. Black holes are the
most exciting gravitational objects and it would be interesting to
study them in pure Lovelock gravity. Black hole uniqueness theorems
do not straightaway carry over in higher dimensions, however they do remain
valid for Einstein gravity so long as horizon has spherical topology \cite{holl}.
We would however stick to spherical horizon topology. Starting from
Einstein-Gauss-Bonnet solution \cite{bd}, static black holes have
been studied quite extensively in Lovelock gravity \cite{whw, btzetc} and
there exists general unique solution for spherical horizon topology
\cite{cai}. Dynamically rotating black hole is the most exciting as
it has rich energetics. Rotating black hole solutions in $5$ dimension have been
studied in Einstein \cite{haw} as well as in Einstein-Gauss-Bonnet gravity \cite{natha, radu}.
It would however be interesting to find a spacetime metric for a rotating
black hole both in Einstein gravity in arbitrary dimension and in pure Lovelock gravity
in even $d=2n+2$ dimension. Note that for pure Lovelock, gravity is kinematic
in all odd $2n+1$ dimensions and so there only exists BTZ black hole. This is why
pure Lovelock rotating black hole metric we wish to obtain cannot be
compared to $5$-dimensional E-GB metric \cite{natha, radu}.

In this letter we wish to obtain rotating black hole metric for both Einstein gravity
in arbitrary higher dimension and pure Lovelock gravity in even $2n+2$ dimension. Instead of solving vacuum equations, which are quite formidable for axially symmetric spacetime, we would  employ two independent methods to obtain metric for rotating black hole
first in Einstein in higher dimension and then in pure Lovelock
gravity. Recently one of us \cite{d-kerr} has employed a novel
method of deriving black hole metric by appealing to the general
principle that a new theory should include existing theory plus the new feature
that necessiated the change. That is Einstein gravity should include Newtonian gravity
In the present context of black hole what it means is that radially falling null and timelike particles respectively experience no acceleration and Newtonian acceleration. It is remarkable
that such simple considerations actually lead to the right metric for both Einstein and pure Lovelock rotating black hole. Second is the ingenious Newman-Janis algorithm (NJA)
\cite{nja} in which one begins with static vacuum solution and
converts it into rotating black hole through a complex
transformation. Interestingly both the methods give the same
metric. It turns out that for Einstein gravity the metric so obtained is indeed
an exact solution of vacuum equation while for pure Lovelock gravity
it satisfies the equation in the leading order. It could be seen as follows.
For pure Lovelock black hole gravitational potential \cite{cai, dpp2} in even $d=2n+2$ dimension
goes as $r^{-1/n}$, Riemann curvature as $r^{-(2n+1)/n}$, then $R^{(n)}_{ab}$
will go as $r^{-(2n+1)}$. The rotating metric we have obtained has
$R^{(n)}_{ab}$ falling off as $r^{-(2n+3)}$, two powers sharper than the expected
leading order. Thus even though (like the famous and useful metric \cite{carmeli}
for a radiating rotating black hole which is not an exact solution for null dust
source in $d=4$) it is not an exact solution of pure Lovelock vacuum
equation yet it is Lovelock Ricci flat in the leading order. It
describes a rotating black hole with all its physical properties like two regular horizons and ergoregion. It is thus a pure Lovelock rotating black hole on the same footing as rotating radiating black hole \cite{carmeli}.

NJA was though invented for constructing a rotating black hole
metric from a static black hole vacuum solution in $4$ dimension,
yet as we would show in the following that it works perfectly well
for all even $d>4$ (even dimension is required for definition
complex tetrad vectors) in Einstein gravity. Not only that a rotating analogue of a
static black hole with Maxwell and Yang-Mills charge could also by obtained \cite{sgg} in all even
dimensions by application of NJA which shows that it is not restricted to vacuum
alone. How it actually works is not
fully understood, however there has been an insightful attempt in this direction \cite{dr-sz}.
That is, on application of NJA to a higher dimensional static black hole
metric \cite{tang}, we do obtain the corresponding rotating black
hole metric which is a solution of Einstein vacuum equation. Despite strong warnings against \cite{patho, 
p-j}, now we wish to extend it to pure Lovelock gravity because it is the most natural extension 
of Einstein gravity. It gives a perfectly well behaved metric for a rotating black hole without any pathologies \cite{patho} and it shares all the properties with Kerr black hole. Generally it is expected that whatever works for Einstein should also work for pure Lovelock. However it is an extrapolation though very strongly motivated. The test of an extrapolation rests on the fact whether it gives
a reasonable and physically acceptable result. That is what indeed it does in the
present context.

\subsection{Incorporating Newtonian gravity} Following Ref. \cite{d-kerr},   we begin with an
ellipsoidal symmetry incorporating rotation about an axis and flat metric for it reads in $d$ dimension as
\begin{eqnarray}
ds^2  & = & \frac{A(r)}{\Sigma}\left(dt - a\sin^2\theta_1
d\theta_2 \right)^2 - \frac{\Sigma}{B(r)}dr^2 - \Sigma d\theta_1^2
- \frac{\sin^2\theta_1}{\Sigma} \nonumber \\
& & \times  \left((r^2+a^2)d\theta_2 - a dt \right)^2 - r^2
\cos^2\theta_1 d\Omega_{(d-4)}^2 \label{eq:flat}
\end{eqnarray}
where $\Sigma = r^2+a^2 \cos^2 \theta_1$ and $A(r) = B(r) =
r^2+a^2$. We can introduce a black hole  into this framework only
through the potential function $A(r)$ because $\Sigma$ defines axial
symmetry which should remain as it is. It can be easily seen that
requirement of radial photon experiencing no acceleration leads to $A(r) =
B(r)$ and let's write it as $A(r)= B(r) = r^2+a^2+\psi$ so that when $\psi=0$,
it is back to flat space. Now asking for incorporation of Newtonian acceleration
$-M/r^{d-2}$, we have $r\psi^{\prime}-2\psi=2r^3(M/r^{d-2})$. This determines $\psi = -2Mr^{(5-d)}$.
This is a rotating black hole metric which is an exact solution of Einstein vacuum equation in
all $d\geq4$ dimensions \cite{haw}. We have the familiar Kerr solution for $d=4$.

To obtain its pure Lovelock analogue we have just to note that
potential in that case \cite{dpp2, cai} goes as
$-(M/r^{d-2n-1})^{1/n}$ and so we have
$r\psi^{\prime}-2\psi=(2/n)(Mr^{4n+1-d})^{1/n}$ which determines
$\psi= -2(M^{1/n}/{d-2n-1})r^{(4n+1-d)/n}$. For $n=1$ it gives the familiar 
Kerr metric which is an axact vacuum solution. For $n>1$, it is not
an exact solution but it is Lovelock Ricci flat in the leading order
for $d=2n+2$ as Ricci curvatures fall off as $r^{-(2n+3)}$, two
orders sharper than the leading order.

The above metric (\ref{eq:flat}) describes an Einstein rotating black hole for $d\geq4$ when

\begin{equation}
 A(r) = B(r) = r^2+a^2-\frac{2M}{r^{d-5}}
\end{equation}

and a pure Lovelock black hole in $d=2n+2$ dimension when

\begin{equation}
 A(r) = B(r) = r^2+a^2-2(M)^{1/n}r^{(2n-1)/n}.
\end{equation}
Clearly in odd $d=2n+1$ dimension potential turns constant indicating kinematic character of pure Lovelock gravity. 
\subsection{Newman-Janis Algorithm}
We begin with static black hole metric which is an exact solution of the corresponding (Einstein or pure Lovelock) vacuum equation with $\Lambda=0$ and it is given by
 \begin{eqnarray}
ds^2 &=&  (1 + \Phi(r))dt^2 - \frac{dr^2}{1+\Phi(r)} + r^2d\Omega_{d-2}^2 \label{solA}
\end{eqnarray}
with
\begin{eqnarray}
d \Omega_{d-2}^2 &=& d \theta^2_{1} + \sin^2({\theta}_1) d
\theta^2_{2} + \sin^2({\theta}_1) sin^2({\theta}_2)d \theta^2_{3}
\nonumber \\ & & + \ldots + \left[\left( \prod_{j=1}^{d-3}
\sin^2({\theta}_j) \right) d \theta^2_{d-2} \right].
\end{eqnarray}
Now
\begin{equation}
 \Phi(r) = -M/r^{d-3}, \, \, \,  -(M/r)^{1/n}
\end{equation}
respectively for Einstein in $d\geq4$ and pure Lovelock in $d=2n+2$ dimension. Let's now apply NJA to the above metric
for which we first transform it to Eddington-Finkelstein coordinate,
$du = dt -  f^{-1}(r) dr$, so that the metric takes the form
\begin{eqnarray}
ds^2 &=&   f(r) du^2 - 2 dudr - r^2 d\Omega_{d-2}^2 \label{SchwEF1}
\end{eqnarray}
where
$$f(r) = 1 + \Phi(r).$$
Following the Newman-Janis prescription \cite{nja, Capo}, we can write the metric in terms of null
veiltrad, $ Z^a = (l^a,\;n^a,\;m_{1}^a,\;\bar{m_{1}}^a,\;m_{2}^a,\;\bar{m_{2}}^a,\;
\ldots, \; m_{k}^a,\;   \bar{m}_{k}^b )$, as
\begin{eqnarray}
{g}^{ab} = l^a n^b + l^b n^a - m_{1}^a \bar{m}_{1}^b - m_{2}^a
\bar{m}_{2}^b,\;\ldots,\;\\ \nonumber -m_{k}^a \bar{m}_{k}^b, \label{NPmetric}
\end{eqnarray}
 where null vieltrad are
\begin{eqnarray*}
l^a &=& \delta^a_r,\\
n^a &=& \left[ \delta^a_u - \frac{1}{2} (1 + \Phi(r))\delta^a_r \right],\\
 m_{k}^a &=& \frac{1}{\sqrt{2}r\sin\theta_{1}\sin\theta_{2}, \;\ldots, \sin\theta_{(k-1)}}  \left( \delta^a_{\theta_{k}}
  + \frac{i}{\sin\theta_{k}} \delta^a_{\theta_{(k+1)}} \right)\\
\end{eqnarray*}
where $k = 1,...,(d-2)/2$. This vieltrad are orthonormal and obey the metric conditions.
\begin{eqnarray}
l_{a}l^{a} = n_{a}n^{a} = ({m_k})_{a} ({m_k})^{a} = (\bar{m_k})_{a} (\bar{m_k})^{a}= 0,  \nonumber \\
l_{a}({m_k})^{a} = l_{a}(\bar{m_k})^{a} = n_{a}({m_k})^{a} = n_{a}(\bar{m_k})^{a}= 0, \; \nonumber \\
l_a n^a = 1, \; ({m_k})_{a} (\bar{m_k})^{a} = 1. \nonumber
\end{eqnarray}
Now we allow for some $r$ factor in the null vectors to take on complex values. We rewrite the null vectors in the form \cite{Capo,Qw}
\begin{eqnarray*}
l^a &=& \delta^a_r, \\
n^a &=& \left[ \delta^a_u - \frac{1}{2} (1 + \Psi(r)) \delta^a_r \right], \\
m_{k}^a &=& \frac{1}{\sqrt{2}\bar{r}\sin\theta_{1}\sin\theta_{2}\; \ldots \;
\sin\theta_{(k-1)}}  \left( \delta^a_{\theta_{k}} + \frac{i}{\sin\theta_{k}} \delta^a_{\theta_{(k+1)}} \right),\\
\end{eqnarray*}
where
\begin{equation}
 \Psi = -\frac{Mr^{d-3}}{r\bar{r}}, \, \, \, -\left(\frac{Mr^{2n-1}}{(r\bar{r})^n}\right)^{1/n}
\end{equation}
respectively for Einstein and pure Lovelock with $\bar{r}$ being complex conjugate of $r$. In $4$-dimension, there is only one axis of rotation and thereby only one angular
momentum parameter. In higher dimensions, there are more than one
axis of rotation and therefore multiple angular momenta. We shall first focus
on the simplest case of only one axis of rotation with angular
 momentum parameter $a$. Further we go

\begin{equation}
{x'}^{a} = x^{a} + ia (\delta_r^{a} - \delta_u^{a}) \cos\theta_1
\rightarrow \\ \left\{\begin{array}{ll}
u' = u - ia\cos\theta_1, \\
r' = r + ia\cos\theta_1, \\
\theta_k' = \theta_k \\
\theta_{k+1}' = \theta_{k+1} \end{array}\right.
\end{equation}
Simultaneously let null vieltrad vectors $Z^a$ undergo a
transformation $Z^a = Z'^a{\partial x'^a}/{\partial x^b} $ and in the
usual way, we obtain
\begin{eqnarray*}
l^a &=& \delta^a_r, \\
n^a &=& \left[ \delta^a_u - \frac{1}{2} (1 + \Psi_1(r) \delta^a_r \right], \\
 m_{k}^a &=& \frac{1}{\sqrt{2}(r+ia\cos\theta_1)\sin\theta_{1}\sin\theta_{2},\; \ldots \;, \sin\theta_{(k-1)}} \nonumber  \\ &&
   \times \left(ia(\delta^a_u-\delta^a_r)\sin\theta_1 + \delta^a_{\theta_{k}} + \frac{i}{\sin\theta_{k}} \delta^a_{\theta_{(k+1)}} \right),\\
 \end{eqnarray*}
where
\begin{equation}
 \Psi_1(r) = -\frac{Mr^{d-3}}{(d-3)\Sigma}, \, \, \, -\left(\frac{Mr^{2n-1}}{\Sigma^n}\right)^{1/n}
\end{equation}
respectively for Einstein and pure Lovelock. Here
$\Sigma$ is as defined earlier and  we have dropped primes.
From this null vieltrad, a new metric is constructed which could, with
somewhat lengthy but straightforward calculations, ultimately be put in
Boyer-Lindquist coordinates to read as
\begin{eqnarray}
ds^2  & = & \frac{\Delta}{\Sigma}\left(dt - a\sin^2\theta_1
d\theta_2 \right)^2 - \frac{\Sigma}{\Delta}dr^2 - \Sigma d\theta_1^2
- \frac{\sin^2\theta_1}{\Sigma} \nonumber \\
& & \times  \left((r^2+a^2)d\theta_2 - a dt \right)^2 - r^2
\cos^2\theta_1 d\Omega_{(d-4)}^2, \label{eq:mtc}
\end{eqnarray}
where
\begin{equation}
\Delta =r^2+a^2- 2Mr^{5-d}, \, \, \, r^2+a^2- 2M^{1/n}r^{(2n-1)/n}
\end{equation}
respectively for Einstein black hole in $d\geq4$ and pure Lovelock black hole in $d=2n+2$. 
This is the same metric as obtained earlier.

Thus we have obtained rotating analogue of the static black hole
metric  (\ref{solA}) in $d\geq4$. This is an exact vacuum solution for
Einstein black hole while a leading order solution for pure Lovelock black hole in $d=2n+2$.   However pure Lovelock rotating black hole metric is an exact vacuum solution for $n=1$ and for
$n>1$, $R^{(n)}_{ab}$ fall off as $r^{-(2n+3)}$, two orders sharper than the leading order.

Eq. (\ref{eq:mtc}) has parameters $M$ and $a$ which are respectively
related to mass $(\mathcal{M})$ and angular momentum $(J)$ via
relations: $\mathcal{M} = \frac{n}{4\pi} A_{2n}M, \;\; J =
\frac{1}{4\pi}A_{2n}M a$ and $\frac{\mathcal{M}}{J} = na$. Area
$A_{2n}$ of unit $2n$-sphere is given by
\begin{eqnarray}
A_{2n}=\int_{0}^{2\pi}d\theta_2 \int_0^\pi \sin\theta_1
\cos^{(2n-2)}\theta_1 d\theta_1  \\ \nonumber \times
\huge\prod_{i=3}^{(2n-2)}\int_{0}^{\pi}\sin^{(2n-2)-i}\theta_{i}d\theta_{i}
 = \frac{2\pi^{(2n+1)/2}}{\Gamma(2n+1)/2}. \label{a}
\end{eqnarray}
where $\sqrt{-g} = \sqrt{\gamma} \Sigma r^{(2n-2)}
\sin\theta\cos^{(2n-2)}\theta$ and $\gamma$ is determinant of $d
\Omega_{(2n-2)}$ metric.

One of the characteristics of rotating black hole is its ring singularity at
$r=0, \theta_1=\pi/2$. Look at the above expressions for $\Delta/\Sigma$ which
is ring singular for Einstein black hole only for $d\leq5$ else it has central singularity. It also turns out that $\Delta=0$ has only one positive root indicating spacelike nature of singularity.  This is quite in contrast to Kerr rotating black hole. On the other hand pure Lovelock black hole is always ring singular with two horizons. It is perfectly in tune with all the 
properties of a rotating black hole. Ergoregion enclosed between static limit ($g_{tt}=0$) and outer horizon (bigger root of $\Delta=0$) however exists for both Einstein and pure Lovelock black holes. 

In a straightforward manner, it is possible to generalize the
rotating black hole metric to include multiple rotation axes and
parameters, e.g., we can easily include two rotation axes with rotation
parameters $a$ and $b$ and in that case the metric would read as
follows:
\begin{eqnarray}
& & ds^2 = \frac{\Delta}{\Sigma} \left(dt - a\sin^2\theta_1 d
\theta_2 - b\cos^2 \theta d \theta_3  \right)^2
- \frac{\Sigma}{\Delta}dr^2 - \Sigma d\theta_1^2 \nonumber \\
& & - \frac{\sin^2\theta_1}{\Sigma} \left((r^2+a^2) \theta_2 - a dt \right)^2 - \frac{\cos^2\theta}{\Sigma}\left(r^2+b^2)d\theta_3 - b dt\right)^2 \nonumber\\
& & - r^2 (\cos^2 \theta_1 + \sin^2\theta_2) d\Omega_{(d-4)}^2,
\label{eq:mtc2}
\end{eqnarray}
where
\begin{eqnarray}
\Delta &=& \frac{(r^2+a^2)(r^2+b^2)}{r^2} + \psi
\label{De}
\end{eqnarray}
where
\begin{equation}
 \psi = -\frac{2M}{r^{d-5}},~~~~ -2M^{1/n} r^{(2n-1)/n}
\end{equation}
respectively for Einstein and pure Lovelock black hole and $\Sigma = r^2 + a^2\cos^2\theta_1 + b^2\sin^2\theta_2$. It will have much richer energetics like that of an Einstein rotating
charged black hole in $5$-dimensional supergravity \cite{kpd}. It would be interesting
to explore it in a future publication.

Even though pure Lovelock black hole metric is not an exact but a leading order solution of Lovelock vacuum equation, yet it describes a rotating black hole wonderfully well with all its expected and desired properties in terms of two regular horizons, ring singularity and ergoregion. These are characteristic properties of rotating black hole. In contrast, Einstein black hole metric  is an exact solution but it has only one horizon thereby singularity is like Schwarzschild black hole spacelike. There is no ring singularity for $d>5$. It does however have an ergoregion. On the other hand, pure Lovelock black hole shares all the properties with Kerr black hole though it satisfies the Lovelock equation in the leading order. The situation is exactly similar to radiating rotating black hole black hole metric \cite{carmeli} obtained by the usual method of transforming static metric into Eddington-Finkelstein coordinates and then letting mass of hole become a function of new time coordinate. But it is not an exact solution of radially outflowing radiation though it has all the desirable physical features. 

Very recently there is a good bit of discussion \cite{patho, p-j} that NJA is inapplicable in non-Einstein theories as it generates naked singularity and pathologies. This may in general be true for other modified gravity theories but certainly not true for pure Lovelock as we have shown that it generates a perfectly well behaved metric for a rotating black hole, though not an exact solution. Pure Lovelock is essentially not a modification of Einstein but it is a general polynomial class of which Einstein is the linear order. However Ref. \cite{patho} brings out a very important point that when higher order terms in curvature are involved, NJA does not necessarily preserve field equation. This may perhaps be the reason for the resulting metric in our case, though bearing all the properties of rotating black hole, is not a solution. It thus  still remains open to find an exact solution and whether it shares all the properties with Kerr black hole? Recall in the case of radially outflowing null radiation from a rotating black hole, there does exist an exact solution but without fathomable physical properties \cite{vp} while one that is not an exact solution possesses all the desirable properties \cite{carmeli}.  

It is interesting to note that very simple common sense considerations in the first method of incorporation of Newtonian gravity lead to rotating black hole metric almost trivially. It is an insightful and novel demonstration of power and elegance of simple minded and physically motivated physical extrapolations. The  hardest thing in there is to find or creatively guess metric imbibing ellipsoidal symmetry, then  the rest is trivial. For Einstein gravity, the method works for all odd and even dimensions. In the case of pure Lovelock, in odd $d=2n+1$ we need $\Lambda$ to give a non-trivial metric. Incorporation of $\Lambda$ in this framework would be quite involved and hence it would be taken up separately later.

Finally we wish to say that we have thus obtained a metric by two well motivated and independent prescriptions and it describes a rotating black hole that shares all the physical properties with the prototype rotating Kerr black hole but it satisfies Lovelock vacuum equation only in leading order. On the other hand we have an exact solution of Einstein vacuum equation but black hole does not share the crucial property of nature of singularity with Kerr black hole. It would be fair to say that pure Lovelock rotating black hole is the only black hole that shares all the properties with Kerr black hole. This is once again realization of universality of pure Lovelock gravity.

\acknowledgements  One of the authors (S.G.G.) would like to thank
University Grant Commission (UGC) for major research project grant
NO. F-39-459/2010(SR)  and to IUCAA, Pune for kind hospitality while
part of this work was being done.


\begin{thebibliography}{99}
\bibitem{dgj}
  N.~Dadhich, S.~G.~Ghosh and S.~Jhingan,
  Phys.\ Lett.\ B {\bf 711}, 196 (2012)
\bibitem{bianchi}
 N. Dadhich, Pramana {\bf74}, 875 (2010) (arXiv:0802.3034)
\bibitem{dpp1}
  N.~Dadhich, J.~M.~Pons and K.~Prabhu,
  Gen.\ Rel.\ Grav.\  {\bf 44}, 2595 (2012) [arxiv:1110.0673]
\bibitem{dpp2}
  N.~Dadhich, J.~M.~Pons and K.~Prabhu,
  Gen.\ Rel.\ Grav.\  {\bf 45}, 1131 (2013)
  [arXiv:1201.4994 [gr-qc]].
 \bibitem{prague}
 N.~Dadhich, The gravitational equation in higher dimensions, Relativity and Gravitation: 100 years after Einstein in Prague, June 25-28, (2012)
  arXiv:1210.3022 [gr-qc].
\bibitem{dgj1}
 N.~Dadhich, S.~G.~Ghosh and S.~Jhingan, Bound orbits and gravitational theory, arxiv:1308.4770v1.
\bibitem{fri}
D. Friedan, Phys. Rev. Lett. {\bf 45}, 1057 (1980); D. J. Gross and E. Witten, Nucl. Phys. {B277}, 1 (1986); M. J. Duff and B. E. W. Nilsson and C. N. Pope, Phys. Lett. {\bf 173}, 69 (1986).
\bibitem{holl}
S. ~Hollands and A. ~Ishibashi, Black hole uniqueness theorems in higher dimensional spacetimes, Topical Invited Review by Class. Quant. Grav., arxiv:1206.1164.
\bibitem{bd}
D.~G. Boulware and S.~Deser, Phys. Rev. Lett. {\bf 55}, 2656 (1985).
\bibitem{whw}
J. ~T. ~Wheeler, Nucl. Phys. {\bf B268}, 727 (1986), {\bf B273}, 732 (1986); B. ~Whitt, Phys. Rev. {\bf D38}, 3000 (1988). 
\bibitem{btzetc}
 M.~Banados, C.~Teitelboim and J.~Zanelli,
  Phys.\ Rev.\ D {\bf 49}, 975 (1994)
\bibitem{cai}
 R.~-G.~Cai and N.~Ohta,
  Phys.\ Rev.\ D {\bf 74}, 064001 (2006), N. Dadhich, Math Today {\bf 26}, 37 (2011)
(arXiv:1006.0337)
\bibitem{haw}
S. ~W. ~Hawking, C. ~J. ~Hunter and M. ~M. ~Taylor-Robinson, Phys. Rev. {\bf59}, 064005 (1999), arxiv:hep-th/9811056.
\bibitem{natha}
A.~Anabalon, N.~Deruelle, Y.~Morisawa, J.~Oliva, M.~Sasaki, D.~Tempo and R.~Troncoso, Class. Quant. Grav. {\bf26}, 065002 (2009) (arxiv:0812.3194); A.~Anabalon, N.~Deruelle, D.~Tempo and R.~Troncoso, Int. J. Mod. Phys. {\bf 20}, 639 (2011) (arxiv:1009.3030).
\bibitem{radu}
Y.~Brihaye, B.~Kleihaus, J.~Kunz and E.~Radu, JHEP {\bf 11}, 098 (2010) (arxiv:1010.0860); B.~Kleihaus, J.~Kunz and E.~Radu, Phys. Rev. Lett. {\bf 106}, 151104 (2011) (arxiv:1101.2868).
\bibitem{d-kerr}
N. ~Dadhich, A novel derivation of the rotating black hole metric, arxiv:1301.5314 (accepeted in Gen. Relativ. Grav.).
 \bibitem{nja}
E. T. ~Newman and A. I. ~Janis, J. Math. Phys. {\bf 6}, 915 (1965); E. T. ~Newman, E. ~Couch, R. ~Chinnapared, A. ~Exton, A. ~Prakash and R. ~Torrence, J. Math. Phys. {\bf 6}, 918 (1965)
\bibitem{carmeli}
M.~Carmeli, Group Theory and General Relativity (Imperial College Press, 2000).
\bibitem{sgg}
  S.~G.~Ghosh and U.~Papnoi,
  arXiv:1309.4231 [gr-qc].
\bibitem{dr-sz}
S. P. Drake and P. Szekeres, An explanation of the Newman-Janis algorithm, arxiv:9807001v1
\bibitem{tang}
F. R. Tengherlini, Nuovo Cimento {bf27}, 636 (1963)
\bibitem{patho}
D. ~Hansen and N. ~Yunes, arxiv:1308.6631 (2013)
\bibitem{p-j}
Y. ~F. ~Pirogov, arxiv:1306.4866 (2013); ~T. ~Johannesen and D. ~Psaltis, arxiv:1304.6631 (2013) 
 \bibitem{btzbh}
M.~Banados, C.~Teitelboim and J.~Zanelli, Phys. Rev. Lett. {\bf 69}, 1849 (1992)
\bibitem{Capo}
 S.~Capozziello, M.~De laurentis and A.~Stabile,  Class.\ Quant.\ Grav.\  {\bf 27}, 165008 (2010)
\bibitem{my} E T.~Newman, R.~Couch, K.~Chinnapared, A.~Exton, A.~Prakash and R.~Torrence,
 J.\ 2Math.\ Phys.\  {\bf 6}, 918 (1965).
\bibitem{Qw} D.~J.~Cirilo Lombardo, Class.\ Quant.\ Grav.\  {\bf 21}, 1407 (2004).
 \bibitem{kpd}
  K.~Prabhu and N.~Dadhich,
  Phys.\ Rev.\ D {\bf 81}, 024011 (2010).

  \bibitem{vp}
 P.~C.~Vaidya and L.~K.~Patel  Phys.\ Rev.\ D {\bf 7}, 3590 (1973).
\end{thebibliography}
\end{document}